\documentclass{icrc29}
\usepackage{graphicx,amssymb,amsmath,times}
\begin{document}
\title{Extragalactic cosmic-ray source composition and the interpretation of the ankle}

\author[D. Allard, E. Parizot, A.V Olinto, E. Khan, S. Goriely]{D. Allard$^{1,2}$, E. Parizot$^3$, A. V. Olinto$^{1,2}$, E. Khan$^3$, S. Goriely$^4$\\(1) Kavli Institute of Cosmological Physics, University of Chicago, 5640 S. Ellis, Chicago, IL60637, USA. 
\\ (2) Department of Astronomy and Astrophysics, University of Chicago, 5640 S. Ellis, Chicago, IL60637, USA. 
\\(3) Institut de Physique Nucl\'eaire d'Orsay, IN2P3-CNRS/Universit\'e Paris-Sud, 91406 Orsay Cedex, France.
\\(4) Institut d'Astronomie et d'Astrophysique, ULB - CP226, 1050 Brussels,
Belgium.} 
\presenter{Presenter: Denis Allard (denis@oddjob.uchicago.edu), \  
usa-Allard-D-abs1-og13-oral}

\maketitle

\begin{abstract}

We consider the stochastic propagation of high-energy protons and nuclei in the cosmological microwave and infrared backgrounds, using revised photonuclear cross-sections and following primary and secondary nuclei in the full 2D nuclear chart. We confirm earlier results showing that the high-energy data can be fit with a pure proton extragalactic cosmic ray (EGCR) component if the source spectrum is $\propto E^{-2.6}$. In this case the ankle in the cosmic ray (CR) spectrum may be interpreted as a pair-production dip associated with the propagation. We show that when heavier nuclei are included in the source with a composition similar to that of Galactic cosmic-rays (GCRs), the pair-production dip is not present unless the proton fraction is higher than 85\%. In the mixed composition case, the ankle recovers the past interpretation as the transition from GCRs to EGCRs and the highest energy data can be explained by a harder source spectrum $\propto E^{-2.2}$-- $E^{-2.3}$, reminiscent of relativistic shock acceleration predictions, and in good agreement with the GCR data at low-energy and holistic scenarios. While the expected cosmogenic neutrino fluxes at high energy are very similar for pure proton and mixed composition hypothesis, the two scenarii predict very different elongation rates from $10^{17.5}$ to $10^{20}$ eV.

\end{abstract}

\section{Introduction}

One of the keys to understanding the origin of cosmic-rays is the spectral shape of the transition from Galactic cosmic-rays  to extragalactic cosmic rays. The Galactic origin of low-energy CRs is generally accepted, while the highest energy CRs  (above $\sim 10^{19}$~eV) are no longer confined by Galactic magnetic fields and most probably originate from other galaxies. Thus, a transition between the two components has to occur in some energy range. The most natural location for this transition is around $\sim 3\,10^{18}$~eV at a feature in the CR spectrum known as the \emph{ankle}. This is the only energy range where the spectrum gets harder (i.e., its logarithmic slope gets smaller), offering a simple transition scheme to a harder CR component which is subdominant at lower energies.

A different conclusion has recently been proposed on the basis of composition results from the HiRes Collaboration, tentatively showing a transition from heavy to light primary nuclei at an energy around $5\,10^{17}$~eV \cite{HiRescomp}, which can be identified with a \emph{second knee} feature in the spectrum. Such composition measurements rely on statistical determinations of the elongation rate of extensive air-showers and depend on shower development simulations which are still quite uncertain.   
In this case the ankle is interpreted as an  e$^{+}$-e$^{-}$ pair-production dip \cite{Berezinsky+02} resulting from proton propagation over large distances in the cosmic microwave background (CMB) and  the transition from GCR to ECGR \emph{does not have} to be at the ankle. For these second-knee-transition models the EGCR component is best fitted by assuming a power-law source spectrum in $E^{-\beta}$ with $\beta \simeq 2.6$ \cite{deMarco+03} or  $\beta \simeq 2.7$ \cite{Berezinsky+02}. These studies, however, assume that EGCRs are only protons.

In this study, we investigate how these conclusions are modified when nuclei heavier than protons are also present in the source. We apply the revised scheme of photo-nuclear interactions described in \cite{Khan+05}. Interestingly, we find that the two main aspects of EGCR phenomenology are jointly affected, namely the best fit slope of the source spectrum and the interpretation of the ankle. The interpretation of the ankle as a pair production dip is not viable in the mixed composition case, unless the fraction of protons is above 85\% . When reasonable assumptions for the fraction of nuclei in the EGCR are introduced, the traditional interpretation of a GCR/EGCR transition at the ankle is recovered and the best fit spectral index for EGCR is  $\sim$ 2.2 -- 2.3, as expected from relativistic shock acceleration. Finally, the best fit EGCR spectrum occurs when both the GCR and EGCR component have the same spectral index, allowing for  holistic approaches to the CR spectrum.

\section{Phenomenological source model}

Although extragalactic magnetic fields could modify the UHECR spectrum, we assume here that their intensity is low enough to be negligible and focus on a comparison between pure proton and mixed-composition models. The choice of a source composition is arbitrary in the absence of a source model, but we consider a ``generic composition'' assuming that EGCRs have the same relative source abundances as the best known, low-energy GCRs, derived from \emph{Ulysses} and \emph{HEAO 3} data \cite{DuvTha96}. We convert the differential abundance ratios, $x_{i}$, at a given energy per nucleon ($E/A$) into ratios at a given energy, $\xi_{i} = x_{i}A_{i}^{\beta-1}$, where $\beta$ is the spectral index, so that the GCR source spectrum is: $N_{i}(E) \propto \xi_{i} E^{-\beta}$. 

Since our goal is also to investigate the viability of the proton pair production dip as an explaination for the ankle when heavier nuclei are emited at the sources we will as well consider proton dominated compositions with small fraction of heavier nuclei. We will study in particular the impact of small fractions of Helium and Iron nuclei. 

Finally, we choose a rigidity dependent maximum energy at the source for the various nuclear species  $E_{\mathrm{max}}(^{A}_{Z}\mathrm{X})$. We assume that energy losses and photo-fragmentation inside the source can be neglected, so that all nuclei with the same gyroradius behave the same way. This implies $E_{\mathrm{max}}(^{A}_{Z}\mathrm{X}) = Z\times E_{\mathrm{max}}(^{1}_{1}\mathrm{H})$.

\section{Results and discussion}

\begin{figure*}[ht]
\centering
\hfill~
\hfill\includegraphics[width=7cm]{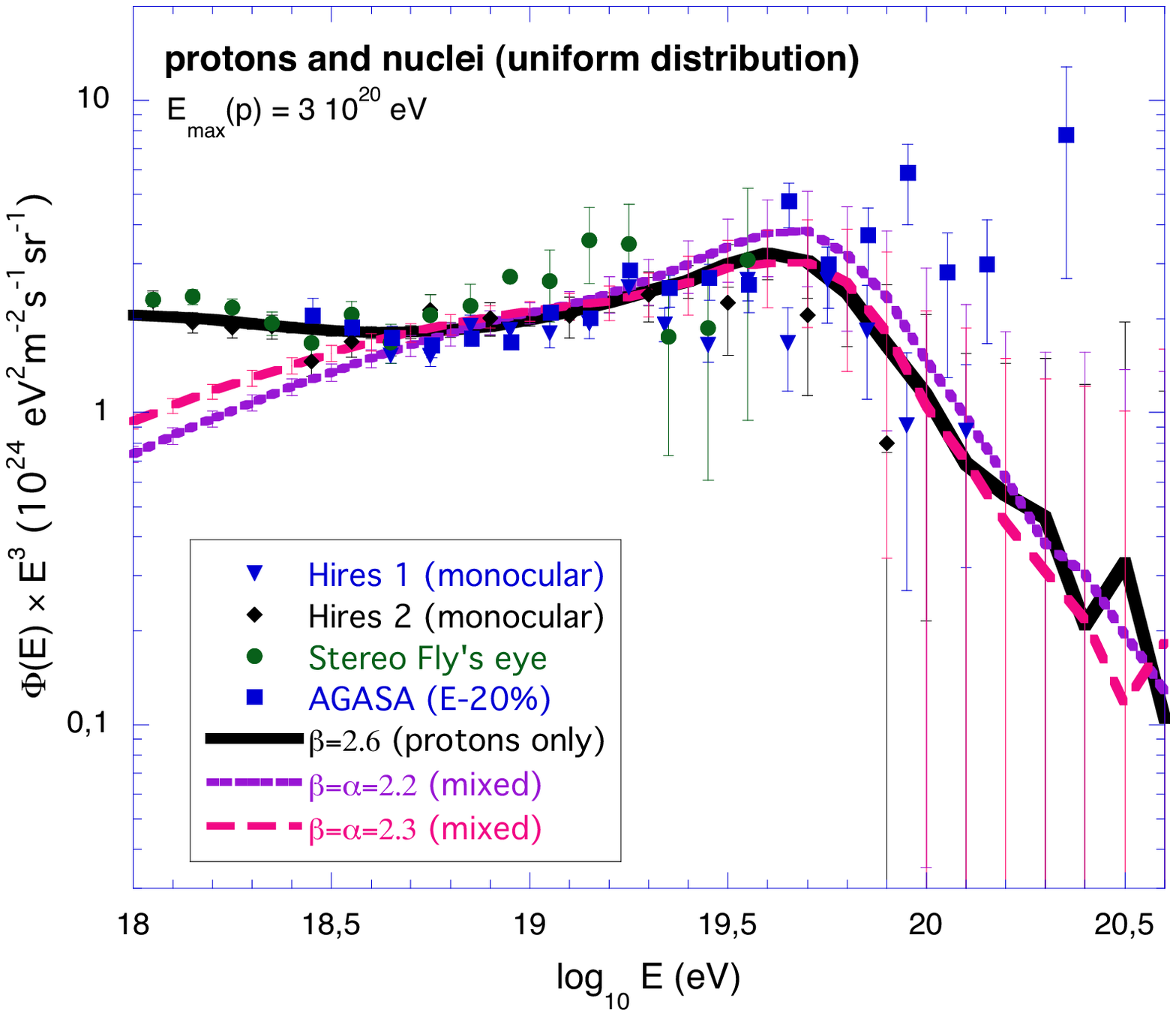}\hfill
\includegraphics[width=7cm]{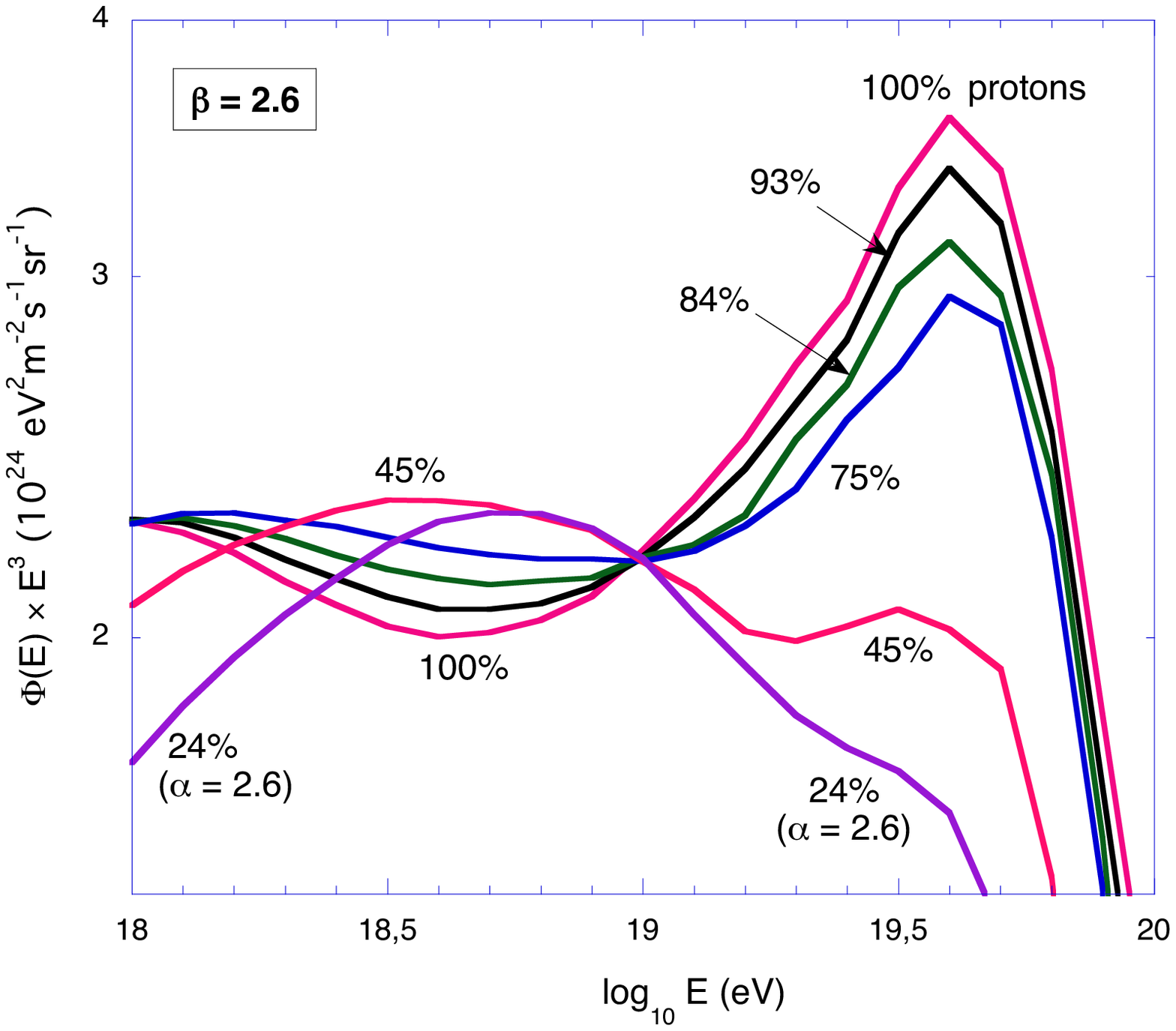}\\\hfill
\includegraphics[width=7cm]{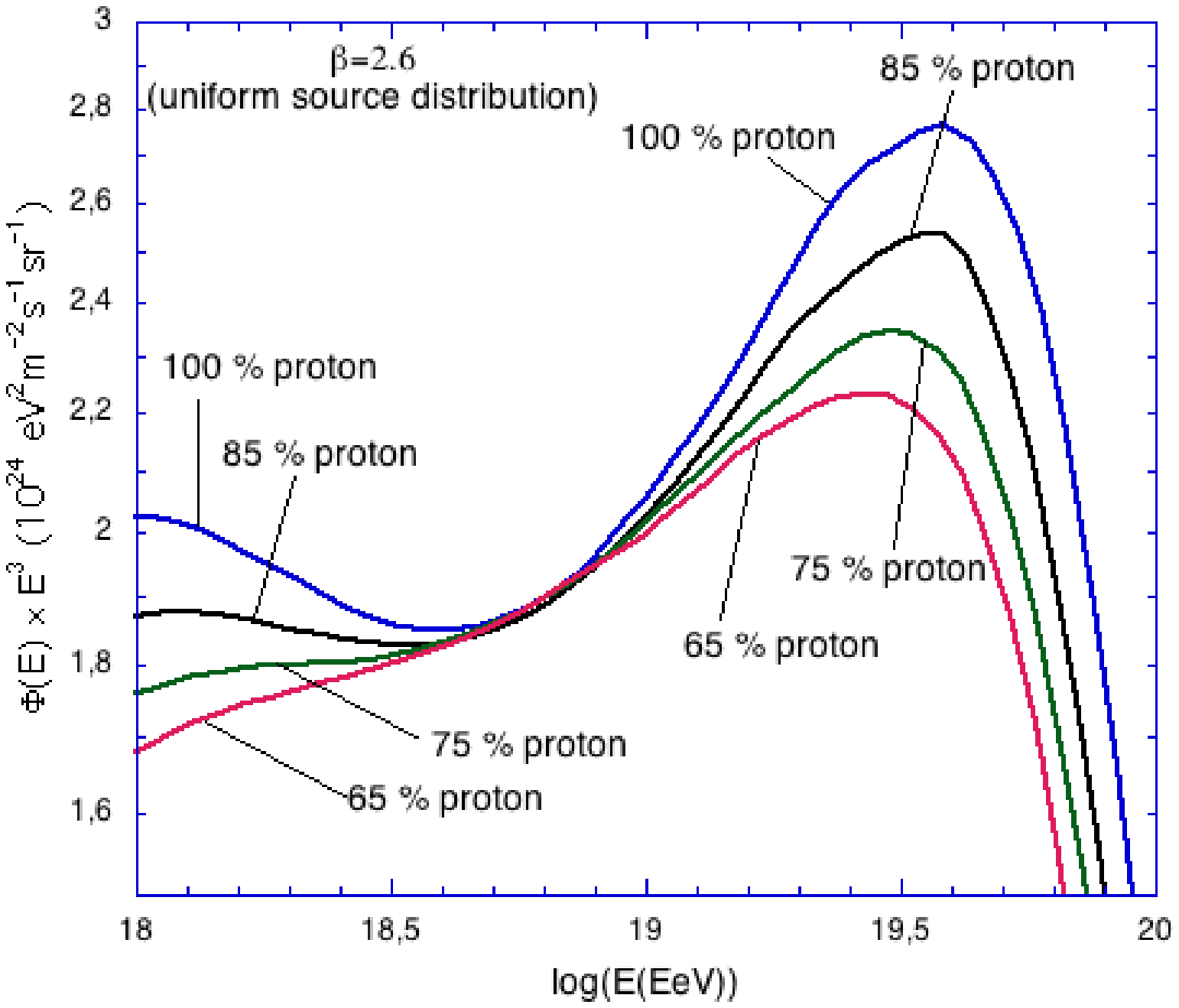}\hfill\includegraphics[width=7cm]{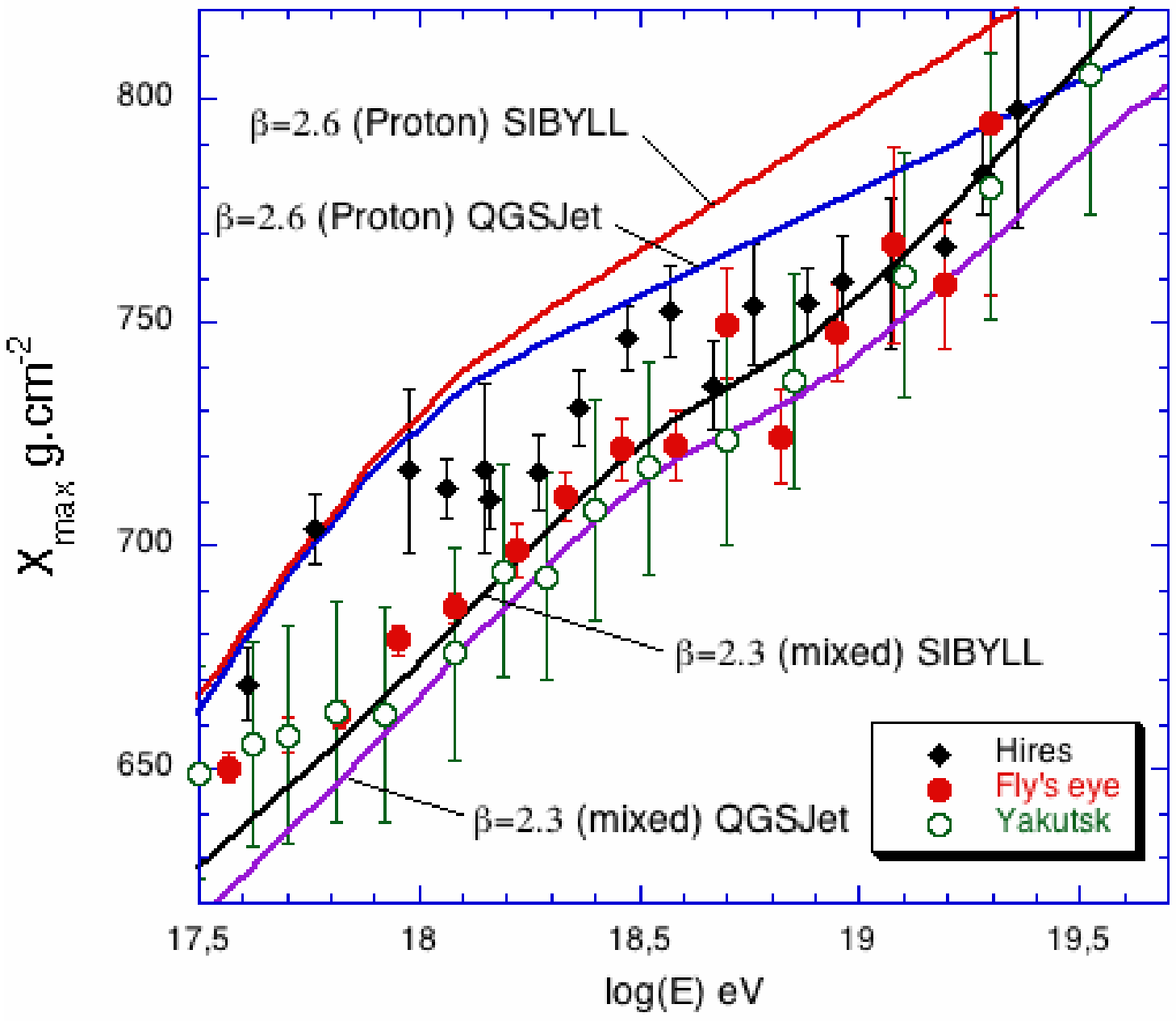}\hfill~
\caption{Top left: Propagated spectra for sources with protons only and $\beta = 2.6$, and with mixed compositions and $\beta = 2.2$ and 2.3. Data points from the main experiments are indicated, with an arbitrary downward shift of AGASA's energy scale by 20\%. Top right: evolution of the spectrum with the fraction of protons at the source, for an $E^{-2.6}$ source spectrum and a heavier compenent dominated by Helium nuclei. Bottom left: Same as top right but with a heavier component dominated by iron nuclei. Bottom right: Elongation Rates corresponding to the pure extragalactic proton and mixed composition cases  for the QGSJet and SIBYLL hadronic models, compared with Hires, Fly's eye and Yakutsk's results.}
\label{fig:spectra}
\end{figure*}
We compute the propagated spectrum of EGCRs for a uniform source distribution with negligible magnetic fields, using a Monte-Carlo technique. For this purpose we take into account all the relevant energy loss processes for protons and nuclei  \footnote{For further details about the treatment of these interactions and further references see \cite{Allard05}.}. 
In Fig.~\ref{fig:spectra}a, we show the resulting propagated spectra obtained with the prescription described above where  $E_{\mathrm{max}}(^{1}_{1}\mathrm{H}) = 3\,10^{20}$~eV.  The error bars represent the 1-$\sigma$ fluctuations of the flux in each energy bin, as expected for a data set with the statistics of the AGASA experiment, i.e., 866 events above $10^{19}$~eV \cite{deMarco+03}. We compare our results with the data from HiRes and AGASA experiments. In the pure proton case, we find again that the best fit spectrum has $\beta = 2.6$ and a pair production dip, as in previous work.  Figures~\ref{fig:spectra}b and~\ref{fig:spectra}c show how the spectrum around the ankle and the GZK bump evolves with the fraction of protons in the source composition (above $10^{18}$~eV) for two composition hypothesis. When the remaining composition is dominated by Helium, for fractions lower than 75\%, the pair production dip is strongly attenuated and the ankle is actually reversed for fractions lower than $\sim 60$\%. When small fraction of iron nuclei are present, the proton pair production dip is also modified but in a different way (due to the difference in the energy losses between He and Fe \cite{Allard05b}) the extragalactic spectrum appears flatter below $5\,10^{18}$~eV and a galactic component is then needed to reproduce the low energy part of the ankle. Therefore, the interpretation of the ankle as an e$^{+}$-e$^{-}$ pair production dip requires a very large fraction of protons at the source. In contrast, Fig.~\ref{fig:spectra}a shows two good fits to the highest energy data using mixed composition models with $\beta = 2.2$ or 2.3, corresponding to 50\% or 40\% protons at the source, respectively. 

The pure proton model with $\beta = 2.6$ provides a good fit to the data down to $\sim 10^{18}$~eV (assuming a uniform source distribution and no magnetic field), the transition from GCR to EGCR should occur before the ankle. However, shock acceleration processes are known to yield a spectrum with  $\beta \simeq 2.2$--2.3. As shown in Fig.~\ref{fig:spectra}a, such spectral indexes can provide equally good fits to the high energy data, assuming a realistic source model with a mixed composition. In such a scenario, the transition from GCRs to EGCRs occurs at the ankle, which thus keeps its ``standard'' interpretation. In addition, injection spectrum $\sim E^{-2.3}$  has also been considered as the best fit to the low-energy data for the GCR component (e.g. \cite{Ptuskin97,StrMos01}). It is thus particularly interesting to note that assuming a similar composition for both the GCR and EGCR components is precisely what also makes a similar source spectrum possible. In addition, it has been shown in \cite{Parizot05} that a source power-law index of $\sim 2.3$ is a necessary condition for holistic models, in which the same sources produce the CRs at all energies. The results presented here show that a self-consistent model can be built for CRs with a similar composition and spectrum at all energies provided the GCR/EGCR transition is at the ankle.

Furthermore, we also calculated the expected cosmogenic neutrino fluxes for the pure proton and mixed model \cite{Allard05b}. For both cases we found very similar high energy neutrino peaks. This can be understood easily since protons dominate the high energy spectrum in the mixed case. We can then conclude for both composition hypothesis high energy neutrino fluxes should be detectable with experiments like Auger or Anita, but these fluxes could certainly not be used to distinguish between the two models. 

In Fig.~\ref{fig:spectra}d, we show the predicted elongation rates for the composition hypothesis of Fig.~\ref{fig:spectra}a, according to 10000 Aires \cite{Sciutto01} simulated showers of the most abundant species \cite{Allard05b} (i.e, H, He, CNO, Mg, Si, Fe) between $10^{17.5}$ and $10^{20}$ eV. For this purpose we assume that the galactic iron component above $10^{17.5}$ eV is the difference between the predicted extragalactic component and the experimental data. In the case of the pure proton extragalactic component, a very steep elongation rate is expected between  $10^{17.5}$ and $10^{18}$ eV in very good agreement with Hires data \cite{HiRescomp}. As it is shown on Fig.~\ref{fig:spectra}b and \ref{fig:spectra}c, since the proton fraction as to be very high in the case of a second knee transition model (i.e a dip due to pair production), the elongation rate is expected to a have the characteristic shape of a rapid transition from a heavy galactic composition (shown by Kascade results \cite{Kascade05}) to a very light extragalactic component unavoidably close to a pure proton expectation. Above $10^{18}$ eV the expected elongation depend strongly on the hadronic models, the consistency of Hires data with this scenario is thus diffucult to establish, one can at least notice that the pure proton case is within the $\sim 25$~g.cm$^{-2}$ HiRes systematic error and the transition is in the expected energy range. For the mixed composition a continuous transition, from galactic iron to  mixed composition (with a flattening between $10^{18.5}$ and $\sim10^{19.2}$ eV) and finally above $10^{19.5}$ eV almost pure proton, is expected. This predicted shape is in good agreement (especially for the SIBYLL model) with Fly's eye and also Yakutsk results \cite{watson00} except below  $10^{18}$ eV where the exact composition of the galactic component becomes relevant. 

Although the predictions are hadronic model dependant, the expected elongation rates appear extremely different whatever the detail of the astrophysical hypothesis \cite{Allard05b} for the two transition models (i.e second knee or ankle). The current disagreements between different experimental results do not allow a conclusion  at this point since each model can be favored or as well almost ruled out, whether one trusts Hires or Fly's eye elongation rate. However,the Pierre Auger Observatory should distinguish between the two hypothesis with  high statistics and high accuracy hybrid elongation rate and finally explain the origin of the ankle.

\end{document}